\documentclass[twoside]{article}
\usepackage{graphicx}
\usepackage{subcaption}
\usepackage{float}
\usepackage{tabularx}
\usepackage[preprint]{aistats2026}
\usepackage[round]{natbib}

\begin{document}

%

%

\twocolumn[

\aistatstitle{Does FOMC Tone Really Matter? Statistical Evidence from Spectral Graph Network Analysis}

\aistatsauthor{Jaeho Choi \And Jaewon Kim \And  Seyoung Chung  \And Chaeshick Chung \And Yoonsoo Lee}

\aistatsaddress{Sogang University}]

\begin{abstract}
    
This study examines the relationship between Federal Open Market Committee (FOMC) 
announcements and financial market network structure through spectral graph theory. 
Using hypergraph networks constructed from S\&P 100 stocks around FOMC announcement 
dates (2011-2024), we employ the Fiedler value—the second eigenvalue of the hypergraph Laplacian—to measure changes in market connectivity and systemic stability. Our event study methodology reveals that FOMC announcements significantly alter network structure across multiple time horizons. Analysis of policy tone, classified using natural language processing, reveals heterogeneous effects: hawkish announcements induce network fragmentation at short horizons ($k=6$) followed by reconsolidation at medium horizons ($k=14$), while neutral statements show limited immediate impact but exhibit delayed fragmentation. These findings suggest that monetary policy communication affects market architecture through a network structural transmission, with effects varying by announcement timing and policy stance.
\end{abstract}

\section{INTRODUCTION}

Monetary policy communications reshape financial markets, yet how they affect market-wide structural cohesion remains underexplored in the literature. This study addresses this gap by measuring changes in network connectivity around FOMC announcements. While prior research has extensively documented the effects of central bank statements on individual asset prices, evidence on their influence over the structure of market relationships is limited.

Our key contribution is introducing the Fiedler value—the second smallest eigenvalue of the hypergraph Laplacian—as a statistical measure of market connectivity that captures system-wide stability. We construct sector-aware hypergraph networks from S\&P 100 stocks and employ an event study framework to measure the effects of FOMC announcements on network structure. Our analysis distinguishes between the immediate impact of information disclosure and the subsequent effects of policy tone. By measuring connectivity changes before and after policy announcements, we quantify how monetary policy reshapes market structure beyond traditional price effects.

Methodologically, we combine three innovations novel in computational finance. First, dual-threshold network construction differentiates within-sector from cross-sector correlations, reflecting heterogeneous relationship formation. Second, hypergraph representation via clique detection captures higher-order interactions among asset groups beyond pairwise connections. Third, residualizing returns on market factors isolates firm-specific co-movements from systematic effects. 

We classify FOMC statements as hawkish, dovish, or neutral using natural language processing, enabling analysis of heterogeneous policy impacts. Our framework captures both the immediate market response to FOMC announcements and the delayed adjustments that materialize as markets gradually process and reassess the information content.

Our analysis highlights a two-stage transmission mechanism. First, the mere act of public disclosure at FOMC dates triggers an immediate re-wiring of market connectivity (the announcement effect). Second, as market participants process the substantive content and follow-on communications, heterogeneous tone effects (hawkish, dovish, neutral) emerge with delays across horizon $k$. This design allows us to separate timing (announcement versus delayed interpretation) from content (policy stance), providing a network-structural perspective on monetary policy communication that complements traditional asset-return evidence.


\section{RELATED WORK}

\subsection{FOMC Announcements and Market Response}
The literature on FOMC announcement effects is extensive, with seminal contributions from \citep{rosa2013financial} demonstrating significant increases in asset price volatility following policy communications. Recent work by \citep{cieslak2019stock} shows that FOMC statements affect not only interest-sensitive assets but also equity markets through expectation channels.

However, these studies predominantly focus on individual asset returns or volatility measures
rather than systemic network effects. The communication aspect of monetary policy has gained prominence, especially since Bernanke’s tenure emphasized forward guidance as a policy tool. Studies by \citep{lucca2015pre} and \citep{cieslak2019stock} document the "pre-FOMC announcement drift" phenomenon, suggesting that markets often anticipate policy decisions before the official release. Beyond immediate price drifts, researchers have begun examining the content of FOMC communications. For example, \citep{hansen2016shocking} use computational
linguistics to quantify the information in FOMC statements and find that the language and tone of announcements can have distinct market impacts beyond the rate decision. 

This aligns with a broader literature recognizing that central bank communication can strongly
affect financial markets. Yet, what remains underexplored is how such communication influences the structure of market relationships. Our research extends this literature by examining whether anticipation effects and communication tone are reflected in network structural changes around FOMC
events.

\subsection{Financial Networks and Systemic Risk}
The application of network theory to finance has grown substantially since the 2008 financial crisis. \citep{acemoglu2016networks} demonstrate how network structure affects systemic risk transmission, showing that the architecture of interconnections can either buffer shocks or amplify contagion. \citep{billio2012econometric} find that financial institutions have become highly interrelated over time, likely increasing the level of systemic risk in the finance and insurance sectors. This underscores that interconnectedness is a double-edged sword: while it may improve risk-sharing under normal conditions, it can facilitate rapid transmission of distress during crises. Recent work by \citep{diebold2014network} develops connectedness measures based on variance decompositions, providing a quantitative framework for understanding spillover effects among financial firms. These studies establish that network topology and
connectivity are critical for assessing systemic stability. 

Within this realm, the use of spectral graph theory in finance is emerging. \citep{sandoval2012cambios} apply correlation-based networks to stock market data, showing that network metrics can offer insights during crisis periods (e.g., identifying market clusters and leadership structure). More recently, \citep{uddin2021attention} constructs a network factor using signed graph Laplacians for asset pricing applications, highlighting that network characteristics can carry price-relevant information. Our research builds on this foundation by focusing specifically on the Fiedler value as a measure of network connectivity and stability. The Fiedler value (second smallest Laplacian eigenvalue) serves as an index of how tightly the market is connected: a larger Fiedler value implies a more cohesive network, whereas a smaller value suggests the network is closer to being disconnected. By examining changes in this measure, we gauge how FOMC announcements shift the cohesion of the financial system.


\section{DATA}

\subsection{FOMC Announcement Data}

Our analysis covers FOMC announcements from 2011 through 2024 (72 scheduled meetings plus emergency announcements). We record each statement of FOMC. FOMC statement tone is classified as hawkish/dovish/neutral using the \textbf{gpt-4o-mini} model. 

We define policy tones as follows: \textit{Hawkish} statements signal 
tighter monetary policy or upward bias in future interest rates, emphasizing 
inflation concerns or economic strength. \textit{Dovish} statements indicate 
accommodative policy or downward rate bias, stressing employment concerns or 
economic risks. \textit{Neutral} statements present balanced risk assessments 
without clear directional bias.

Our dataset comprises 72 FOMC meetings, with 29 classified as hawkish (40.3\%), 25 as dovish (34.7\%), and 18 as neutral (25.0\%). Examples: December 2015 (rate liftoff) is hawkish, March 2020 (150 bps emergency cut) is dovish, and rate holds with balanced risk language are neutral. This tone classification is used in heterogeneity analysis.

\subsection{Stock Market Data}

The analysis uses S\&P 100 daily closing prices from 2000 to 2025 (100 large-cap stocks). To remove market-wide effects, each stock's returns are regressed on SPY and QQQ, and residuals are used for subsequent analysis.

Network construction incorporates Global Industry Classification Sector(GICS) information through a dual-threshold scheme: intra-sector connections require a lower correlation threshold, while inter-sector connections require a higher threshold. This captures both within-industry cohesion and cross-industry linkages.

Event dates are drawn from scheduled FOMC meetings between 2011 and 2024, serving as reference points for constructing symmetric pre- and post-event windows in the event study analysis.

\subsection{Control Variables}

To isolate the effect of FOMC announcements on network connectivity, we include several financial market controls in all baseline regressions.

\subsubsection{Market Volatility}
We control for the level and daily change of the CBOE Volatility Index (VIX), which capture the prevailing uncertainty in equity markets as well as any contemporaneous shocks to volatility on the announcement day.

\subsubsection{Market Return}
We include the daily return of the S\&P 500 index. Large aggregate price movements may influence cross-stock correlations through common shocks, independent of monetary policy announcements.

\subsubsection{Interest Rates}
Both the 2-year and 10-year U.S. Treasury yields are added as controls. These yields serve as benchmarks for short- and long-term interest rate expectations, providing a proxy for broader macro-financial conditions that could affect equity market connectivity.

\subsubsection{Exchange Rate}
We also control for the U.S. Dollar Trade-Weighted Index (TWI), as exchange rate fluctuations can reflect shifts in global risk sentiment and capital flows that may influence stock market comovements.





By including these controls, we isolate the Fed announcement effect from confounders. Results will be reported both with and without the full control set to demonstrate robustness.


\section{METHODOLOGY}

\subsection{Network Construction}

We construct stock market networks by employing a symmetric $k$-window algorithm. The sample consists of daily prices of S\&P 100 constituents, and every trading day is treated as a reference point. For each reference day $t$, the pre-event and post-event windows are defined as follows:

\begin{align}
\text{Pre-window} &= [t - k, t - 1] \\
\text{Post-window} &= [t + 1, t + k]
\end{align}

where $k$ ranges from 5 to 20 trading days, allowing the analysis to capture both short- and medium-term dynamics. All windows are automatically filtered to ensure a minimum number of valid observations, and missing values or extreme returns (e.g., above $\pm$50\%) are either removed or winsorized to avoid distortions.

Daily stock returns are computed as following equation:
\begin{equation}
r_t = \frac{P_t - P_{t-1}}{P_{t-1}}
\end{equation}

To control for market-wide effects, each stock's returns are regressed on SPY and QQQ ETF returns, and the residuals are used for the subsequent correlation analysis \citep{wu2015review}. Pearson correlation matrices are then computed from these residual returns, ensuring that the resulting networks capture firm-specific co-movements rather than aggregate market dynamics.

In constructing the networks, sector classification information is incorporated. Each stock is mapped to its respective GICS sector, and different correlation thresholds are applied: a lower threshold is used for intra-sector connections, while a higher threshold is imposed on inter-sector connections. This dual-threshold design allows the networks to reflect both industry-level cohesion and cross-industry interactions.

The correlation matrices are subsequently transformed into adjacency matrices $A$ by thresholding. Hyperedges are generated by applying a maximum clique detection algorithm. Hyperedge sizes are restricted to between 3 and 12 nodes, with at most 400 cliques retained per window. A computational time limit of 0.75 seconds per window is imposed to ensure efficiency, and a fallback mode retains only 3-cliques in the event of excessive density. These restrictions are introduced to prevent overly dense hypergraphs that reduce economic interpretability and to ensure computational tractability across all event windows.

Finally, for each window, the networks are evaluated using the Fiedler value ($\lambda_2$) and the corresponding Fiedler vector, which capture network connectivity and latent partition structures. For the hypergraphs, the Zhou Laplacian is additionally employed to compute $\lambda_2$, providing a measure of collective connectivity.

\subsection{Hypergraph Connectivity and Eigenvalue Analysis}

We evaluate network connectivity primarily on \emph{hypergraphs}. Let $H$ denote the vertex--edge incidence matrix, $D_v$ the diagonal matrix of vertex degrees, and $D_e$ the diagonal matrix of hyperedge cardinalities. Following \citep{zhou2006learning}, the normalized hypergraph Laplacian is
\[
L_H \;=\; I \;-\; D_v^{-1/2}\, H \, D_e^{-1} \, H^\top \, D_v^{-1/2}.
\]
For an undirected hypergraph, $L_H$ is symmetric and positive semi-definite with eigenvalues
$0=\lambda_1^H \le \lambda_2^H \le \cdots \le \lambda_n^H$.
The second-smallest eigenvalue, $\lambda_2^H$, is our primary measure of \emph{algebraic connectivity}: larger values indicate stronger group-level cohesion, whereas smaller values indicate fragmentation.

Hyperedges are constructed from \emph{residual} correlation networks (stock returns residualized on SPY and QQQ) using a sector-aware thresholding rule (lower threshold within sector, higher across sectors), after which maximal cliques (size 3--12) are extracted. This design captures higher-order co-movements beyond pairwise links. The associated second eigenvector (the ``hypergraph Fiedler vector'') provides a data-driven bipartition that we examine around FOMC announcements to assess sectoral alignments.

\subsection{Event Study Framework}
To demonstrate the effect of FOMC event on network connectivity, we defined $\Delta Fiedler$ as follows :
\[\Delta Fiedler_t = Fiedler_{t,post} - Fiedler_{t,pre}, \; \forall t \in \mathcal{T}\] 
,where  $\mathcal{T}$ is a set of time index in our dataset. Thus, we can measure the change of network connectivity each date. Moreover, we defined $FOMC_t$ dummy variable as follows: 
\[FOMC_t = \left\{\begin{array}{ll} 
     1   \quad \text{if}  \; t \in\mathcal{E}\\
     0  \quad \text{Otherwise}
\end{array} \right.\]
Here, $\mathcal{E}$ denotes the set of all dates which FOMC event held. $X_t$ is a vector of control variables at time $t\in \mathcal{T}$. 

Our baseline regression model is specified as follows:

\[ \Delta Fiedler_t = \beta_0 + \beta_F FOMC_t+  \gamma^{\prime}X_t +\varepsilon_t \]

A critical methodological challenge in this specification is the potential for effect contamination, as FOMC meetings occasionally occur in close temporal proximity. The post-event window of one announcement could overlap with the pre-event window of a subsequent announcement. This overlap would create a methodological contagion, where the lingering effect of the first event contaminates the baseline measurement for the second, thereby biasing the estimation of its unique impact.

To ensure the analytical integrity and robustness of our estimates, we implement a strict exclusion criterion. Any observation period where the full event window ($[t-k, t+k]$) of one FOMC meeting overlaps with that of another is explicitly removed from our regression sample. This procedure ensures that each event's estimated impact is isolated and independent of confounding effects from adjacent policy announcements.

Within this robust framework, the coefficient of interest, \textbf{$\beta_F$}, quantifies the average marginal effect of an FOMC announcement on the change in network connectivity. It represents the estimated shift in $\Delta \text{Fiedler}$ on announcement days relative to non-announcement days, after controlling for prevailing market conditions.

\subsection{Heterogeneous Treatment Effects}
The baseline model in the previous section estimates the average effect of an FOMC announcement. However, the theoretical background suggests that the market's reaction may vary significantly depending on the policy tone conveyed. To investigate these heterogeneous effects, we disaggregate the $FOMC_t$ dummy into three distinct categories based on our NLP classification: \textbf{Hawkish, Dovish, and Neutral.}

We define three mutually exclusive dummy variables. Let $\mathcal{E}_H, \mathcal{E}_D, \mathcal{E}_N$ be the sets of dates for Hawkish, Dovish, and Neutral FOMC announcements, respectively, such that $\mathcal{E} = \mathcal{E}_H \cup \mathcal{E}_D \cup \mathcal{E}_N \; \text{and} \; \mathcal{E}_H \cap \mathcal{E}_D \cap \mathcal{E}_N = \emptyset$. The dummy variables are then defined as follows:
\begin{itemize}
    \item $Hawkish_t = 1$ if $t \in \mathcal{E}_H$, and 0 otherwise.
    \item $Dovish_t = 1$ \quad if $t \in \mathcal{E}_D$, and 0 otherwise.
    \item $Neutral_t = 1$ \; if $t \in \mathcal{E}_N$, and 0 otherwise.
\end{itemize}

By substituting these variables into our baseline model, we can estimate the distinct impact of each policy tone. The model is specified as:
\begin{align*}
    \Delta Fiedler_t = \beta_0 &+ \beta_H \cdot Hawkish_t + \beta_D \cdot Dovish_t \\
    & + \beta_N \cdot Neutral_t + \gamma^{\prime} X_t + \varepsilon_t
\end{align*}

In this specification, the coefficients $\beta_H, \beta_D,$ and $\beta_N$ capture the additional change in network connectivity for each respective tone, compared to the baseline of a non-FOMC day ($\beta_0$).


\section{RESULTS}
\subsection{Baseline Effect}

FOMC announcements have a significant average effect on the Fiedler value of stock market networks. In other words, $\beta$ in our baseline model will be non-zero, indicating that the act of the Fed making an announcement (regardless of content) tends to move the network connectivity away from its pre-event baseline.

\begin{figure}[h]
    \centering
    \vspace{.3in}
        \includegraphics[width=\linewidth]{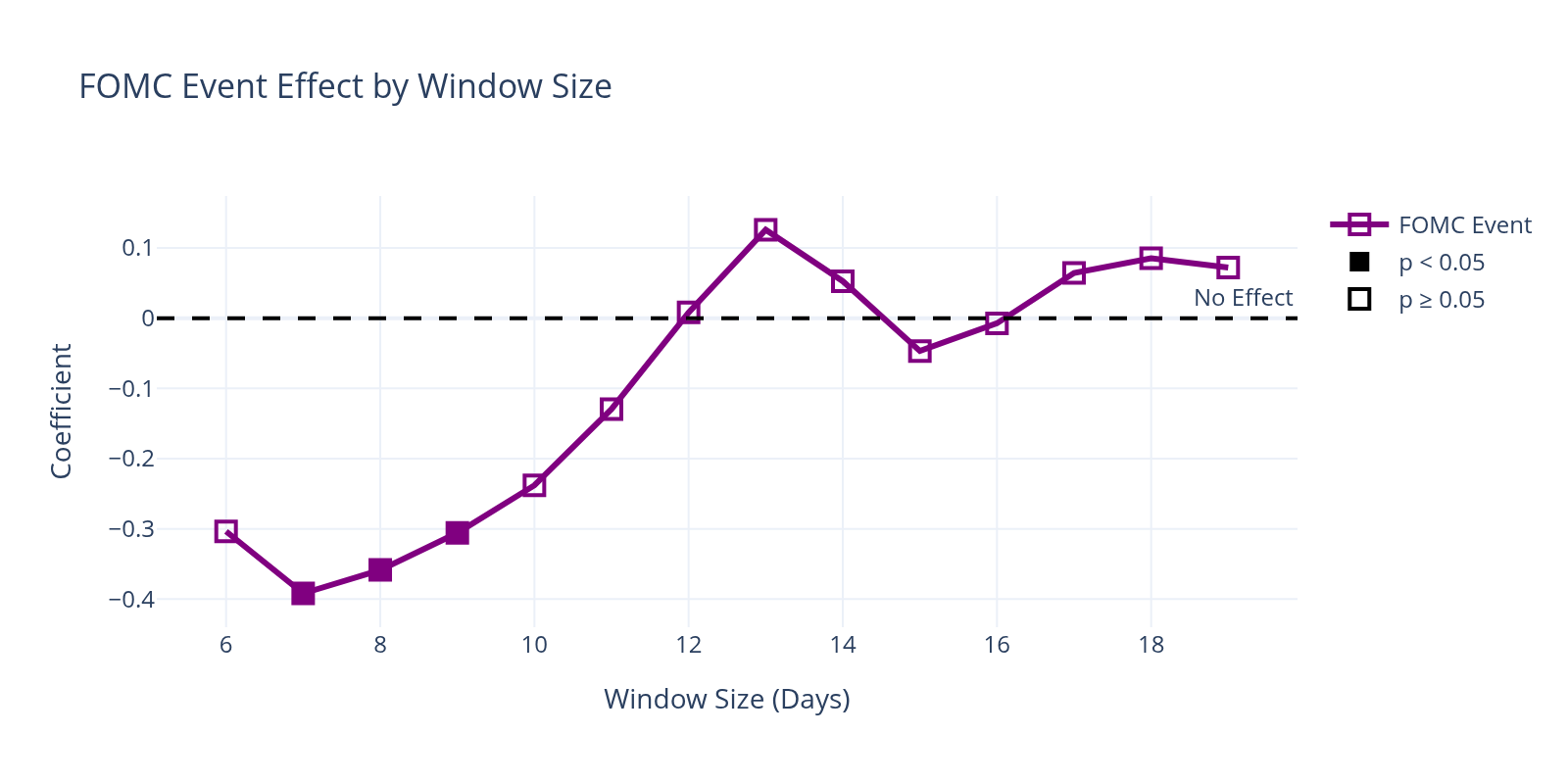}
    \vspace{.3in}
    \caption{Estimated FOMC announcement effects across different event windows. Black squares indicate statistical significance at the 5\% level}
    \label{fig:two_panels}
\end{figure}

We test whether the act of communicating policy—irrespective of content—systematically shifts market connectivity. we find that $\beta_F \neq 0$ for a non-trivial subset of event windows. Significance clusters appear at at a mid horizon ($k=7$–$9$ days). Coefficients also exhibit a sign reversal over $k$: negative at very short horizons, turning positive at intermediate horizons. This dynamic adjustment indicates that FOMC announcements are associated with evolving, rather than instantaneous, shifts in network connectivity.(See Figure~\ref{fig:two_panels}) 

Model fit is modest but consistent with event-study expectations: out-of-sample variance explained ranges from $0.1\%$ to $0.6\%$ across window sizes (see Figure~\ref{fig:r_sq}). These results reject the null of no announcement effect and establish that FOMC communications, as such, induce statistically detectable re-wiring of equity market networks.

These baseline results indicate that FOMC announcements are associated with statistically detectable shifts in network connectivity. 
Importantly, the magnitude and significance of these coefficients are robust to 
alternative specifications without the full set of controls, suggesting that the findings are not sensitive to control selection.

\begin{table}[h]
\caption{Baseline FOMC Effects on Network Connectivity} 
\label{tab:model2-fomc}
\begin{center}
    
\small
\begin{tabular}{@{}l*{4}{c}@{}}
\hline\hline
\textbf{Variable} & \textbf{k=7} & \textbf{k=8} & \textbf{k=9} & \textbf{k=14} \\
\hline
FOMC       & $-0.392^{**}$  & $-0.359^{**}$  & $-0.306^{**}$  & $0.053$       \\
           & (0.015)        & (0.061)        & (0.089)        & (0.753)       \\
\hline
F-statistic     & 2.647      & 1.284      & 1.322      & 0.316      \\
P-value         & 0.015      & 0.261      & 0.244      & 0.929      \\
$R^2$           & 0.006      & 0.004      & 0.003      & 0.001      \\
Adj. $R^2$      & 0.004      & 0.001      & 0.001      & $-0.002$   \\
AIC             & 9,647      & 7,043      & 7,707      & 4,663      \\
BIC             & 9,688      & 7,083      & 7,748      & 4,702      \\
Observations    & 2,666      & 2,128      & 2,399      & 1,984      \\
\hline\hline
\multicolumn{5}{@{}p{\columnwidth}@{}}{\footnotesize Notes: $^{***}$p$<$0.01, $^{**}$p$<$0.05, $^{*}$p$<$0.1. Dependent variable: $\Delta$Fiedler value. Controls include VIX, S\&P 500 return, 2Y/10Y Treasury yields, and Dollar TWI. Results are robust to alternative specifications without these controls, yielding qualitatively similar coefficients.} \\
\end{tabular}
\end{center}
\end{table}

\subsection{Heterogenous Effect}

The impact on network connectivity varies with contextual factors: policy stance, communication form, and economic conditions.

\begin{figure}[h]
    \centering
    \vspace{.3in}
   \includegraphics[width=\linewidth]{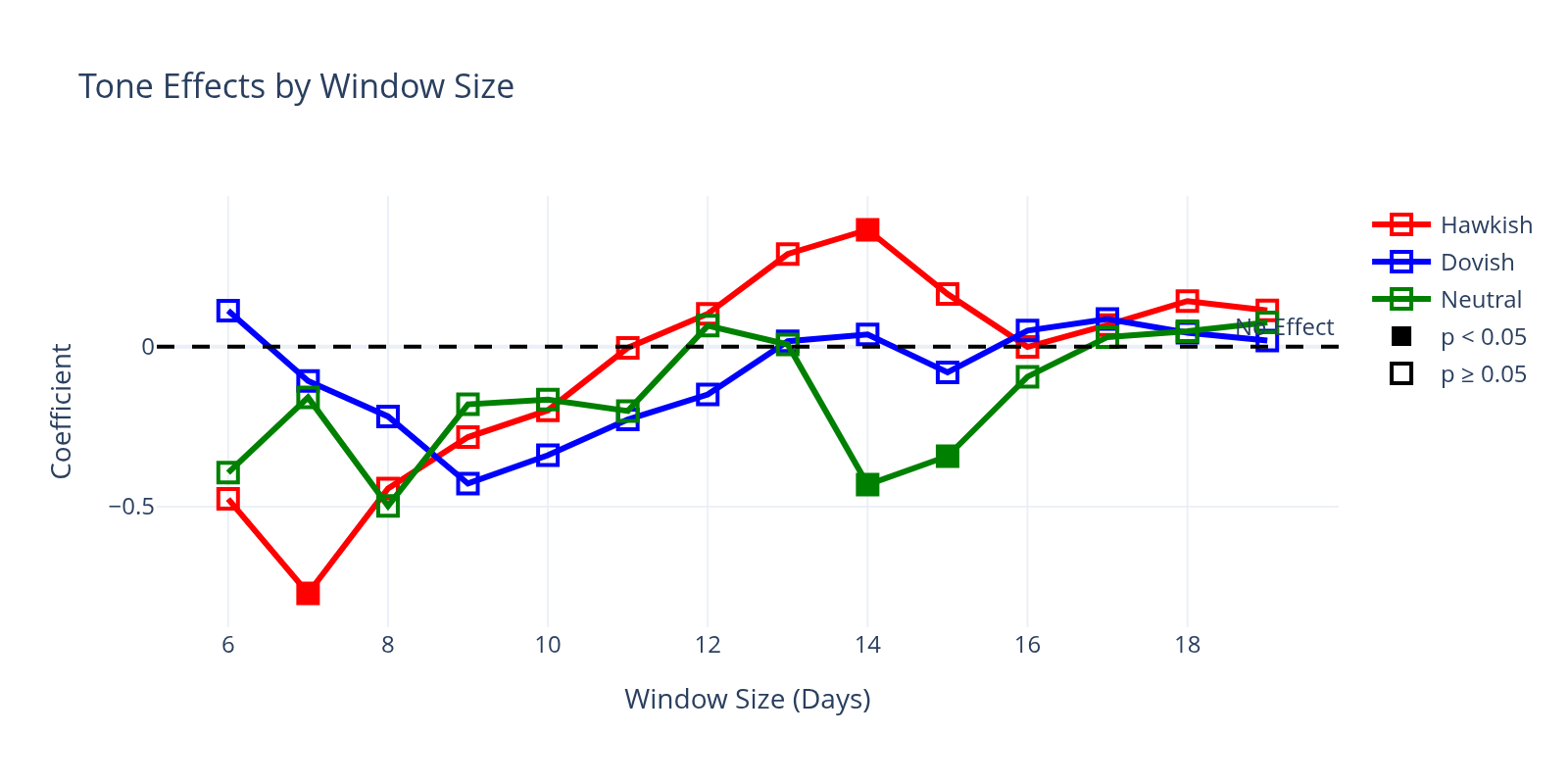}
    \vspace{.3in}
    \caption{Estimated heterogeneous tone effects across different event windows. Black squares indicate statistical significance at the 5\% level}
    \label{fig:hetero_panels}
\end{figure}

The results reveal a systematic heterogeneity across tone categories(See Figure ~\ref{fig:hetero_panels}):
\begin{itemize}

  \item \textbf{Hawkish:} Negative coefficients at short horizons ($k=6$) indicate fragmentation of the network (lower $\Delta \lambda_2$), consistent with heterogeneous sectoral reactions to tightening. However, $k=14$, it strengthens the connection of the network.
  
  \item \textbf{Dovish:} Dovish tone effects are not statistically significant for all time window $k$. 
  
  \item \textbf{Neutral:} Immediate effects are muted, but significant negative coefficients emerge at mid horizons ($k=14$–$15$), consistent with delayed fragmentation. A plausible interpretation is that initially “balanced” language is subsequently re-read as absence of guidance, amplifying uncertainty and belief dispersion. While we treat this interpretation as hypothesis rather than identification, the mid-horizon association is robust across specifications. 
\end{itemize}

\begin{figure}[h]
    \centering
    \vspace{.3in}
    \includegraphics[width=\linewidth]{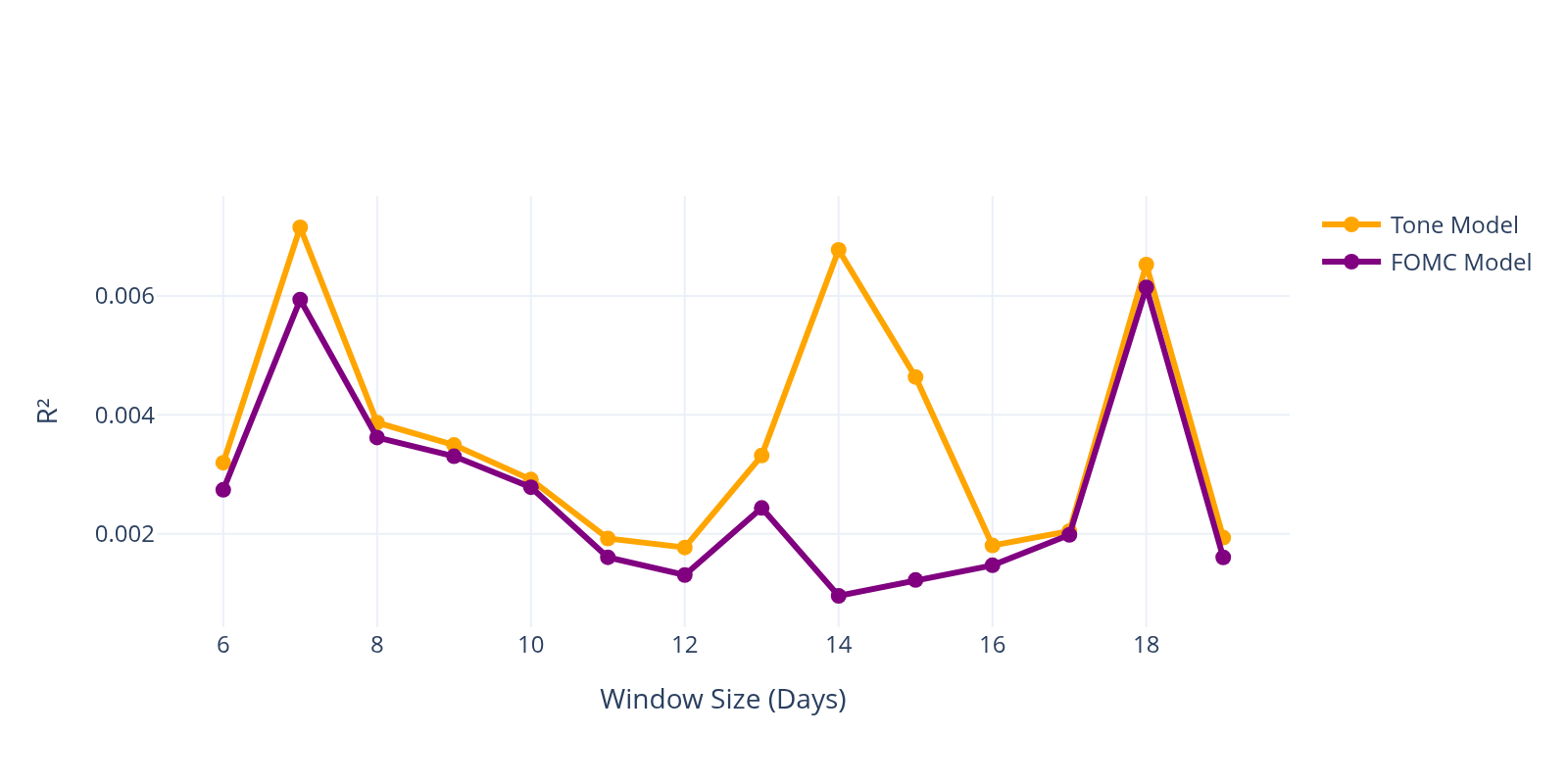}
    \vspace{.3in}
    \caption{$R^2$ score with time window $k$.}
    \label{fig:r_sq}
\end{figure}

 Across windows, tone-decomposed specification outperforms the baseline model in terms of $R^2$ and frequency of significant coefficients (see Figure~\ref{fig:r_sq}), demonstrating that tone and context materially enhance explanatory power beyond the mere occurrence of announcements.

The heterogeneous tone effects remain qualitatively similar when controls are dropped one by one or jointly, indicating that the documented differences across hawkish, dovish, and neutral communications are robust to control choice.

\begin{table}[h]
\caption{Heterogeneous Effects of FOMC Tone on Network Connectivity} 
\label{tab:model1-tone}
\begin{center}
   
\small
\begin{tabular}{@{}l*{4}{c}@{}}
\hline\hline
\textbf{Variable} & \textbf{k=7} & \textbf{k=8} & \textbf{k=9} & \textbf{k=14} \\
\hline
Hawkish    & $-0.772^{***}$ & $-0.443$       & $-0.283$       & $0.366^{**}$  \\
           & (0.000)        & (0.056)        & (0.183)        & (0.012)       \\[0.5ex]
Dovish     & $-0.107$       & $-0.218$       & $-0.428^{*}$   & $0.038$       \\
           & (0.643)        & (0.326)        & (0.051)        & (0.847)       \\[0.5ex]
Neutral    & $-0.159$       & $-0.498^{*}$   & $-0.180$       & $-0.432^{**}$ \\
           & (0.505)        & (0.029)        & (0.420)        & (0.028)       \\
\hline
F-statistic     & 2.392      & 1.029          & 1.047          & 1.684         \\
P-value         & 0.014      & 0.411          & 0.398          & 0.098         \\
$R^2$           & 0.007      & 0.004          & 0.003          & 0.007         \\
Adj. $R^2$      & 0.004      & 0.000          & 0.000          & 0.003         \\
Observations    & 2,666      & 2,128          & 2,399          & 1,984         \\
\hline\hline
\multicolumn{5}{@{}p{\columnwidth}@{}}{\footnotesize Notes: $^{***}$p$<$0.01, $^{**}$p$<$0.05, $^{*}$p$<$0.1. Dependent variable: $\Delta$Fiedler value. Controls include VIX, S\&P 500 return, 2Y/10Y Treasury yields, and Dollar TWI. Results are robust to alternative specifications without these controls, yielding qualitatively similar coefficients.} \\
\end{tabular}
\end{center}
\end{table}

\section{CONCLUSION}

This paper provides evidence that the transmission of monetary policy extends beyond its well-documented effects on asset prices to systematically reconfigure the structure of financial market networks. The substantive content—or tone—of the announcement and the temporal horizon over which its effects are measured are critical determinants of this structural impact. Using spectral analysis on sector-aware hypergraphs of S\&P 100 constituents, we provide evidence consistent with a network-based channel of monetary policy communication.

Our central findings reveal a dynamic and heterogeneous adjustment process. FOMC announcements induce structural shifts in market connectivity that vary by policy tone and time horizon. Hawkish announcements induce short-run fragmentation ($\lambda_2$ decreases), suggesting divergent sectoral responses to anticipated policy tightening. Dovish communications show positive effects on market cohesion ($\lambda_2$ increases), though these effects are more modest and less consistent across horizons than hawkish fragmentation. Neutral statements exhibit a delayed effect, leading to significant network fragmentation at medium horizons. This underscores a crucial insight: analyzing effects across multiple time windows reveals heterogeneity that a single-window approach would miss.

These results carry three broader implications for economic analysis. First, provide evidence for a network-based mechanism for the transmission of monetary policy, where central bank communication alters the covariance structure and potential for contagion within the financial system. Second, the findings offer a quantitative framework for assessing systemic connectivity shifts. The changes in algebraic connectivity we document provide measures of network structural adjustments in response to policy signals. Third, our methodology can be adapted to analyze the structural impact of information shocks in other high-dimensional economic systems where agent inter-dependencies are paramount.

Future research could extend this framework to other major central banks, alternative asset classes, and higher-frequency intraday data to further map the intricate relationship between public information and the evolving topology of financial markets. Ultimately, our findings argue for a re-conceptualization of policy influence: central bank communications do not merely inform market participants; they reshape the structure of their inter-relationships.

\clearpage
\appendix
\thispagestyle{empty}



\end{document}